\documentclass{ceurart-arxiv}

\usepackage{booktabs}
\usepackage{listings}
\usepackage{enumitem}
\usepackage{url}
\usepackage{array}
\usepackage{tabularx}
\usepackage{graphicx}
\usepackage{xcolor}
\usepackage{tikz}
\usepackage{placeins}
\usepackage[ruled,vlined]{algorithm2e}

\lstdefinelanguage{turtle}{
  morekeywords={@prefix,a},
  sensitive=true,
  morecomment=[l]{\#},
  morestring=[b]",
}

\lstdefinestyle{rdfstyle}{
  basicstyle=\ttfamily\small,
  breaklines=true,
  columns=fullflexible,
  keepspaces=true,
  showstringspaces=false,
  frame=single
}

\newcommand{\stepbadge}[1]{%
  \tikz[baseline=(step.base)]{%
    \node[fill=black,text=white,circle,minimum size=1.35em,inner sep=0pt,font=\scriptsize\bfseries] (step) {#1};%
  }%
}

\begin{document}

\copyrightyear{2026}
\copyrightclause{Copyright for this paper by its authors.
Use permitted under Creative Commons License Attribution 4.0 International (CC BY 4.0).}

\conference{DMKG’26: 2nd International Workshop on Data Management for Knowledge Graphs, October 2026, Bari, Italy}

\title{GQL Datatypes as an RDF Vocabulary}

\author[1]{Dominik Tomaszuk}[email=dominik.tomaszuk@uwb.edu.pl]
\author[1]{Daniel Szałaj}
\address[1]{Faculty of Computer Science, University of Bialystok, Poland}

\begin{abstract}
The ISO GQL standard introduces a rich value-type system for property graphs, including scalar, temporal, byte-string, list, record, and dynamic union datatypes. The vocabulary presented here additionally covers a vector extension. At the same time, RDF-based knowledge graph systems identify literal datatypes by IRIs and interpret typed literals through value spaces, lexical spaces, and lexical-to-value mappings. This paper presents the \emph{GQL Datatypes RDF Vocabulary}, a lightweight RDF vocabulary that represents selected GQL datatypes and the vector extension as RDF datatype IRIs and describes their hierarchy, aliases, and selected constraints using RDF, RDFS, OWL, and XML Schema Datatypes vocabulary terms. The vocabulary is intended to support interoperability between GQL-oriented property graph systems and RDF-based knowledge graph systems. We describe its namespace and datatype model, summarize its datatype coverage, and show how the accompanying generator and validator use the vocabulary. In the reference implementation, RDF literals retain GQL datatype IRIs; XSD mappings declared in the vocabulary support scalar validation, while lists, records, and vectors are encoded directly as JSON lexical forms of vocabulary-typed literals.
\end{abstract}


\maketitle

\section{Introduction}
\label{sec:introduction}

Knowledge graph systems increasingly need to exchange data across heterogeneous graph models. RDF systems represent information as triples or datasets and identify literal datatypes by IRIs. Property graph systems, in contrast, usually expose values through a host-language or database-language type system in which node and edge properties may contain booleans, strings, numbers, dates, lists, records, vectors, or other structured values. The standardization of GQL as a database language for property graphs makes this type-system gap more visible: GQL provides a common language-level vocabulary for property graph values, whereas RDF provides a common Web-level mechanism for identifying and interpreting typed literals~\cite{iso-gql,rdf12-concepts}.

This gap is important for data management for knowledge graphs. Many workflows combine RDF-based KGs with property graph systems, for example, when RDF graphs are loaded into graph databases, when property graph data is exported as RDF, when graph embeddings are attached to KG entities, or when validation and schema artifacts are exchanged between graph ecosystems. In these settings, datatype information is often either lost, flattened into generic strings, or approximated by existing XML Schema datatypes. Such approximations are sufficient for common scalar values, but they are less satisfactory for GQL-specific or constructed values such as lists, records, vectors, dynamic union values, and implementation-defined numeric ranges.

This paper presents a short description of the \emph{GQL Datatypes RDF Vocabulary}. The vocabulary assigns RDF datatype IRIs to selected GQL value datatypes and records datatype relationships and constraints using standard RDF vocabulary terms. It is not a replacement for the GQL language specification and does not define a complete GQL parser. Instead, it provides a machine-readable bridge between the GQL value-type vocabulary and RDF's typed literal mechanism. As such, it can support graph model interoperability, datatype-aware validation, schema exchange, and documentation of datatype assumptions in hybrid RDF/property-graph data management pipelines.

The contribution of this paper is threefold:
\begin{enumerate}[leftmargin=*]
  \item we describe a compact RDF representation of GQL datatype names as datatype IRIs;
  \item we outline a datatype model based on value spaces, lexical spaces, and lexical-to-value mappings, aligned with RDF and XML Schema Datatypes;
  \item we present a validation-oriented representation strategy for GQL-typed RDF literals, using ontology-derived XSD mappings for scalar checks and direct JSON lexical forms for constructed datatypes.
\end{enumerate}

The remainder covers the motivation (Section~\ref{sec:motivation}),
vocabulary design (Section~\ref{sec:vocabulary-design}), datatype coverage
and mapping (Section~\ref{sec:coverage-mapping}), implementation and
evaluation (Section~\ref{sec:implementation-evaluation}), related work
(Section~\ref{sec:related-work}), and conclusions
(Section~\ref{sec:conclusion}).

\section{Motivation}
\label{sec:motivation}

RDF provides a general mechanism for typed literals. A literal such as
\texttt{"42"\textasciicircum\textasciicircum xsd:integer} is interpreted using the datatype IRI, its lexical space, its value space, and the lexical-to-value mapping defined for that datatype. This design is extensible: applications may introduce additional datatype IRIs, provided that their interpretation is specified.

However, datatype interoperability between RDF and property graphs remains incomplete. Many GQL scalar values can be mapped naturally to existing XSD datatypes, for example, booleans to \texttt{xsd:boolean}, signed integers to \texttt{xsd:integer}, unsigned integers to \texttt{xsd:nonNegativeInteger}, strings to \texttt{xsd:string}, and dates to \texttt{xsd:date}. Other GQL values have no complete XSD counterpart, including dynamic union values, records, lists, vectors, and approximate numeric values whose precision and range may not be preserved by narrowing them to \texttt{xsd:float} or \texttt{xsd:double}.

The issue is also practical: datatypes influence validation, query evaluation, indexing, storage layout, and data quality. In large-scale KGs, inconsistent handling of datatypes may lead to inefficient queries, failed joins, or subtle semantic errors. A shared datatype vocabulary can help systems document and exchange datatype assumptions, for example, by preserving that a source value was a \texttt{gql:INT8} even if the materialized RDF literal is represented as \texttt{xsd:integer}.

Recent work on composite RDF literals for lists and maps argues that modeling composite values using RDF collections or containers is verbose and difficult to query and update in SPARQL, and proposes dedicated literal datatypes for compact list and map values~\cite{hartig2024datatypes,sparqlcdts}. The GQL Datatypes RDF Vocabulary follows a similar intuition, but with a different scope: it provides an RDF vocabulary for identifying and constraining GQL value datatypes and validating RDF literals that retain those datatype IRIs.

\section{Vocabulary Design}
\label{sec:vocabulary-design}

\subsection{Namespace and Core Terms}

The prototype vocabulary uses the namespace \url{http://example.org/gql-types#} with the preferred prefix \texttt{gql}. Each datatype represented by the vocabulary is identified by an IRI in this namespace. For example, \texttt{gql:BOOLEAN}, \texttt{gql:INT64}, \texttt{gql:STRING}, \texttt{gql:DATE}, \texttt{gql:LIST}, \texttt{gql:RECORD}, and \texttt{gql:VECTOR} are datatype IRIs.

The vocabulary uses \texttt{rdfs:Datatype} to type every datatype IRI, \texttt{rdfs:subClassOf} to represent direct datatype generalization, \texttt{owl:equivalentClass} to represent aliases or equivalent datatype names, and \texttt{owl:withRestrictions} to encode validator-oriented XSD-style constraints. The ontology also uses \texttt{owl:onDatatype}; these terms are consumed as validation metadata and do not establish complete OWL 2 semantics for GQL. A representative fragment is shown in Listing~\ref{lst:vocab-fragment}.

\begin{lstlisting}[caption={Illustrative fragment of the GQL datatype vocabulary.},label={lst:vocab-fragment}]
@prefix gql:  <http://example.org/gql-types#> .
@prefix rdfs: <http://www.w3.org/2000/01/rdf-schema#> .
@prefix owl:  <http://www.w3.org/2002/07/owl#> .
@prefix xsd:  <http://www.w3.org/2001/XMLSchema#> .

gql:BOOLEAN
    a rdfs:Datatype ;
    rdfs:label "BOOLEAN"@en ;
    rdfs:subClassOf gql:BOOLEAN_DATA ;
    owl:withRestrictions (
        [ xsd:pattern "TRUE|FALSE|true|false|1|0" ]
    ) .

gql:BOOL
    a rdfs:Datatype ;
    rdfs:label "BOOL"@en ;
    rdfs:subClassOf gql:BOOLEAN_DATA ;
    owl:equivalentClass gql:BOOLEAN .
\end{lstlisting}

\subsection{Datatype Model}

Each datatype is described, where applicable, with reference to the usual three-part datatype model:
\begin{description}[leftmargin=*]
  \item[Value space.] The set of abstract values denoted by well-formed literals of the datatype.
  \item[Lexical space.] The set of Unicode strings accepted as lexical forms.
  \item[Lexical-to-value mapping.] The mapping from accepted lexical forms to values in the value space.
\end{description}

The vocabulary records selected validator-oriented constraining facets using OWL and XML Schema Datatypes vocabulary terms. These facets are interpreted conjunctively by the reference validator. Thus, if a datatype is associated with both a pattern and a numeric bound, a valid lexical form must satisfy the pattern and denote a value satisfying the bound. For scalar datatypes, these constraints can often be expressed precisely using XSD facets~\cite{xsd11-datatypes}. For constructed datatypes, such as lists, records, and vectors, the implementation uses JSON lexical forms on literals whose datatype remains the corresponding GQL datatype IRI; the inner parsing is implemented in the validator rather than as RDF or OWL datatype semantics.

For the purposes of this paper, the vocabulary can be described as a tuple
\[
\mathcal{V}=(D,H,A,B,R),
\]
where \(D\) is the finite set of GQL datatype IRIs, \(H \subseteq D \times D\) is the hierarchy relation encoded with \texttt{rdfs:subClassOf}, \(A \subseteq D \times D\) is the alias relation between GQL datatypes, \(B\) is a partial mapping from GQL datatypes to XSD validation bases, and \(R\) assigns lexical or value restrictions to selected datatypes.

This distinction is deliberate. The vocabulary does not attempt to provide a complete RDF datatype semantics for all GQL datatypes. In particular, constructed GQL datatypes such as lists and records, and extension types such as vectors, are treated as vocabulary-typed literals whose lexical forms are validated by the reference validator. Thus, the vocabulary should be understood as a lightweight datatype identification and validation profile rather than as a full formal reconstruction of the GQL value system in RDF semantics.

\section{Datatype Coverage and Mapping}
\label{sec:coverage-mapping}

\subsection{Datatype Coverage}

The current vocabulary contains 87 \texttt{rdfs:Datatype} definitions. They can be grouped into several families: abstract base datatypes, dynamic union datatypes, Boolean datatypes, numeric datatypes, character string datatypes, binary string datatypes, temporal datatypes, list datatypes, record datatypes, and vector datatypes. The vector family is introduced by this prototype to represent numeric embeddings; it is not defined by ISO GQL.

Its coverage can be summarized by the family tuple
\[
\mathcal{C}=(D_{\mathrm{base}},D_{\mathrm{dynamic}},D_{\mathrm{boolean}},D_{\mathrm{numeric}},D_{\mathrm{string}},D_{\mathrm{binary}},D_{\mathrm{temporal}},D_{\mathrm{list}},D_{\mathrm{record}},D_{\mathrm{vector}}),
\]
where each component is a subset of \(D\) and the vocabulary-level coverage is their union. Table~\ref{tab:families} summarizes representative datatype families and their intended RDF role.

\begin{table}
\caption{Representative datatype families in the RDF vocabulary.}
\label{tab:families}
\begin{tabular}{lll}
\toprule
Family & Examples & RDF role \\
\midrule
Boolean & \texttt{gql:BOOLEAN}, \texttt{gql:BOOL} & typed scalar values \\
Integer & \texttt{gql:INT8}, \texttt{gql:INT64}, \texttt{gql:INT256} & exact numeric values \\
Unsigned integer & \texttt{gql:UINT8}, \texttt{gql:UINT64} & non-negative exact values \\
Decimal & \texttt{gql:DECIMAL}, \texttt{gql:DEC} & exact decimal values \\
Approximate numeric & \texttt{gql:FLOAT}, \texttt{gql:DOUBLE} & precision-sensitive values \\
String & \texttt{gql:CHAR}, \texttt{gql:VARCHAR}, \texttt{gql:STRING} & Unicode strings \\
Binary & \texttt{gql:BINARY}, \texttt{gql:VARBINARY}, \texttt{gql:BYTES} & octet sequences \\
Temporal & \texttt{gql:DATE}, \texttt{gql:ZONED\_DATETIME} & temporal values \\
Constructed & \texttt{gql:LIST}, \texttt{gql:RECORD}, \texttt{gql:VECTOR} & structured values \\
Dynamic & \texttt{gql:ANY}, \texttt{gql:ANY\_VALUE} & union-like values \\
\bottomrule
\end{tabular}
\end{table}

The hierarchy is useful for at least two reasons. First, it makes GQL datatype organization available to RDF tools. For instance, a validator can discover that \texttt{gql:INT8} is an exact numeric datatype by following \texttt{rdfs:subClassOf} links. Second, it enables schema-level mappings that preserve more information than direct conversion to generic RDF literals. For example, even if a value is materialized as an \texttt{xsd:integer} literal for compatibility, the source schema may still state that the originating GQL datatype was \texttt{gql:INT8}.

\subsection{Mapping GQL Values to RDF Literals}

The reference implementation does not normalize generated values to XSD literals or \texttt{rdf:JSON} literals. Instead, RDF literals retain GQL datatype IRIs, and the ontology supplies XSD mappings and restrictions used during validation. Constructed values are represented as JSON lexical forms on GQL-typed literals. Table~\ref{tab:mapping} summarizes the validation targets and lexical conventions used by this implementation.

A GQL-typed RDF literal is modeled here as a pair
\[
l=(x,d),
\]
where \(x\) is the lexical form and \(d \in D\) is a GQL datatype IRI. Validation is represented by the partial decision function
\[
\operatorname{valid}(x,d)=
\begin{cases}
\operatorname{jsonShape}(x,d), & d \in D_{\mathrm{list}}\cup D_{\mathrm{record}}\cup D_{\mathrm{vector}},\\
\operatorname{booleanLexical}(x), & d \in D_{\mathrm{boolean}},\\
\operatorname{scalarValid}(x,d,B(d)) \land \operatorname{facets}(x,R(d)), & B(d)\ \text{is defined},\\
\bot, & \text{otherwise.}
\end{cases}
\]
The cases cover JSON structures, case-insensitive Boolean lexical forms, and other scalars with ontology-derived XSD validation bases and restrictions. Here, \(\operatorname{facets}(x,R(d))\) is true when \(x\) satisfies every applicable pattern, length constraint, and numeric bound in \(R(d)\).

\begin{table}
\caption{Representative validation targets for GQL-typed RDF literals.}
\label{tab:mapping}
\small
\renewcommand{\arraystretch}{0.92}
\begin{tabularx}{\linewidth}{@{}>{\raggedright\arraybackslash}p{0.56\linewidth}>{\raggedright\arraybackslash}X@{}}
\toprule
GQL datatype or family & Validator target \\
\midrule
\texttt{gql:BOOLEAN}, \texttt{gql:BOOL}
  & \texttt{xsd:boolean} \\

\texttt{gql:INT8}, \texttt{gql:INT16}, \texttt{gql:INT32}
  & \texttt{xsd:byte}, \texttt{xsd:short}, \texttt{xsd:int} \\

\texttt{gql:INT64}, \texttt{gql:INT128}, \texttt{gql:INT256}
  & \texttt{xsd:long} or \texttt{xsd:integer} \\

\texttt{gql:INTEGER}, \texttt{gql:BIGINT}, \texttt{gql:SMALLINT}
  & \texttt{xsd:integer} \\

\texttt{gql:SIGNED\_INTEGER}, \texttt{gql:SIGNED\_BIG\_INTEGER}
  & \texttt{xsd:integer} \\

\texttt{gql:UINT8}, \texttt{gql:UINT16}, \texttt{gql:UINT32}
  & \texttt{xsd:unsignedByte}, \texttt{xsd:unsignedShort}, \texttt{xsd:unsignedInt} \\

\texttt{gql:UINT64}, \texttt{gql:UINT128}, \texttt{gql:UINT256}
  & \texttt{xsd:unsignedLong} or \texttt{xsd:nonNegativeInteger} \\

\texttt{gql:UNSIGNED\_INTEGER}, \texttt{gql:UBIGINT}, \texttt{gql:USMALLINT}
  & \texttt{xsd:nonNegativeInteger} \\

\texttt{gql:DECIMAL}, \texttt{gql:DEC}
  & \texttt{xsd:decimal} \\

\texttt{gql:FLOAT}, \texttt{gql:REAL}
  & \texttt{xsd:double} \\

\texttt{gql:DOUBLE}, \texttt{gql:DOUBLE\_PRECISION}
  & not mapped in validator \\

\texttt{gql:FLOAT16}, \texttt{gql:FLOAT32}, \texttt{gql:FLOAT64}
  & \texttt{xsd:float} or \texttt{xsd:double} \\

\texttt{gql:FLOAT128}, \texttt{gql:FLOAT256}
  & \texttt{xsd:double} \\

\texttt{gql:CHAR}, \texttt{gql:VARCHAR}, \texttt{gql:STRING}
  & \texttt{xsd:string} \\

\texttt{gql:BINARY}, \texttt{gql:VARBINARY}, \texttt{gql:BYTES}
  & \texttt{xsd:hexBinary} \\

\texttt{gql:DATE}
  & \texttt{xsd:date} \\

\texttt{gql:LOCAL\_TIME}, \texttt{gql:TIME\_WITHOUT\_TIME\_ZONE}
  & \texttt{xsd:time} \\

\texttt{gql:ZONED\_TIME}, \texttt{gql:TIME\_WITH\_TIME\_ZONE}
  & \texttt{xsd:time} \\

\texttt{gql:LOCAL\_DATETIME}, \texttt{gql:TIMESTAMP}
  & \texttt{xsd:dateTime} \\

\texttt{gql:TIMESTAMP\_WITHOUT\_TIME\_ZONE}
  & \texttt{xsd:dateTime} \\

\texttt{gql:ZONED\_DATETIME}, \texttt{gql:TIMESTAMP\_WITH\_TIME\_ZONE}
  & \texttt{xsd:dateTime} \\

\texttt{gql:DURATION\_YEAR\_TO\_MONTH}
  & \texttt{xsd:duration} \\

\texttt{gql:DURATION\_DAY\_TO\_SECOND}
  & \texttt{xsd:duration} \\

\texttt{gql:LIST}, \texttt{gql:ARRAY}, \texttt{gql:GROUP\_LIST}
  & JSON array lexical form \\

\texttt{gql:GROUP\_ARRAY}
  & JSON array lexical form \\

\texttt{gql:RECORD}, \texttt{gql:ANY\_RECORD}, \texttt{gql:UNIT\_RECORD}
  & JSON object lexical form \\

\texttt{gql:VECTOR}
  & JSON numeric array lexical form \\

\bottomrule
\end{tabularx}
\end{table}

The use of JSON is limited to the lexical representation of constructed values~\cite{rfc8259}. JSON was chosen because RDF 1.2 recognizes JSON as a datatype through \texttt{rdf:JSON}~\cite{rdf12-concepts}, making it a natural notation for structured literal values. However, this design deliberately differs from representing such values directly as \texttt{rdf:JSON} literals. The datatype IRI remains, for example, \texttt{gql:ARRAY}, \texttt{gql:VECTOR}, or \texttt{gql:UNIT\_RECORD}, so that the source datatype is preserved; the JSON text itself is the lexical form. The validation procedure is summarized in Algorithm~\ref{alg:gql-to-rdf}.

\begin{algorithm}[t]
\DontPrintSemicolon
\KwIn{A typed RDF literal $l$ and the GQL datatype ontology}
\KwOut{A validation result}
$d \leftarrow \textsc{LocalName}(\textsc{Datatype}(l))$\;
$x \leftarrow \textsc{LexicalForm}(l)$\;
\If{$d$ is a list, record, unit record, or vector datatype}{
  $j \leftarrow \textsc{ParseJSON}(x)$\;
  \Return \textsc{CheckExpectedJSONShape}($j,d$)\;
}
\If{$d$ is \texttt{BOOLEAN} or \texttt{BOOL}}{
  \Return $\textsc{Lowercase}(x) \in \{\texttt{true},\texttt{false},\texttt{1},\texttt{0}\}$\;
}
$t \leftarrow \textsc{OntologyXSDMapping}(d)$\;
\If{$t$ exists}{
  \Return \textsc{CheckScalarLexicalForm}($x,d,t$) $\land$
    \textsc{CheckOntologyFacets}($x,d$)\;
}
\Return \textsc{UnknownDatatypeError}($d$)\;
\caption{Validation of GQL-typed RDF literals in the reference implementation.}
\label{alg:gql-to-rdf}
\end{algorithm}

The validator first handles datatypes whose lexical forms need structural parsing. Lists, arrays, group lists, group arrays, and vectors must parse as JSON arrays; vectors are additionally restricted to numeric entries. Records and any-records must parse as JSON objects, and unit records must parse as an empty JSON object. Booleans use a case-insensitive lexical check. Other scalar datatypes are checked through the ontology: \texttt{owl:equivalentClass} supplies the XSD base datatype, while \texttt{owl:withRestrictions} supplies patterns, numeric bounds, and length constraints. Datatypes that are neither explicitly structural nor mapped to an XSD datatype are reported as unknown by the reference validator.

\subsection{Examples}

Consider a property graph row or property record containing a Boolean flag, a small integer score, a timestamp, a list of tags, and a vector embedding. The RDF fragment in Listing~\ref{lst:gql-values} illustrates how such values can be represented using the GQL datatype IRIs.

The example can be written as a property tuple with an associated datatype tuple:
\[
\begin{aligned}
r &= (\mathit{active},\mathit{score},\mathit{created},\mathit{tags},\mathit{vector}),\\
\tau(r) &= (\text{\texttt{gql:BOOLEAN}},\text{\texttt{gql:INT8}},\text{\texttt{gql:ZONED\_DATETIME}},\\
&\quad \text{\texttt{gql:LIST}},\text{\texttt{gql:VECTOR}}).
\end{aligned}
\]
Listing~\ref{lst:gql-values} materializes this tuple as RDF predicate--object pairs whose literal datatypes are GQL datatype IRIs. Predicates identify the fields; triple order is immaterial. Recovering column order requires an external schema or explicit position metadata. JSON arrays retain their element order within each literal.

\begin{lstlisting}[caption={Typed GQL values represented as RDF literals.},label={lst:gql-values}]
@prefix ex:  <https://example.org/resource/> .
@prefix gql: <http://example.org/gql-types#> .

ex:row1
    ex:active  "true"^^gql:BOOLEAN ;
    ex:score   "42"^^gql:INT8 ;
    ex:created "2026-05-05T12:30:00Z"^^gql:ZONED_DATETIME ;
    ex:tags    "[\"alpha\", \"beta\"]"^^gql:LIST ;
    ex:vector  "[0.125, -2.5, 9.0]"^^gql:VECTOR .
\end{lstlisting}

This representation is useful when preserving the original GQL datatype is more important than immediately normalizing every value to a standard RDF datatype. It may be used by schema exchange tools, RDF exports of property graph data, and validation pipelines that need to retain the source datatype vocabulary. A separate normalization step may still map selected values to XSD or \texttt{rdf:JSON} targets for compatibility with RDF systems that do not recognize the GQL datatype IRIs directly.

\section{Implementation and Evaluation}
\label{sec:implementation-evaluation}

\subsection{Implementation}

The reference implementation is intended to demonstrate and test the proposed GQL datatype vocabulary rather than to introduce a separate data-generation framework. It consists of three main artifacts: the Turtle ontology \texttt{gql\_datatypes.ttl}, the generator \texttt{generator\_real.py}, and the validator \texttt{validator.py}, all available in the public repository.\footnote{\url{https://github.com/szalajdaniel/Data_types_for_GQL_in_RDF_letters}} The ontology declares 87 datatype IRIs, including the vector extension, in the namespace \url{http://example.org/gql-types#}. It records hierarchy, aliases, XSD bases, and selected lexical and value constraints with the RDF and OWL terms described above.

\begin{figure}[htbp]
\centering
\includegraphics[width=\linewidth]{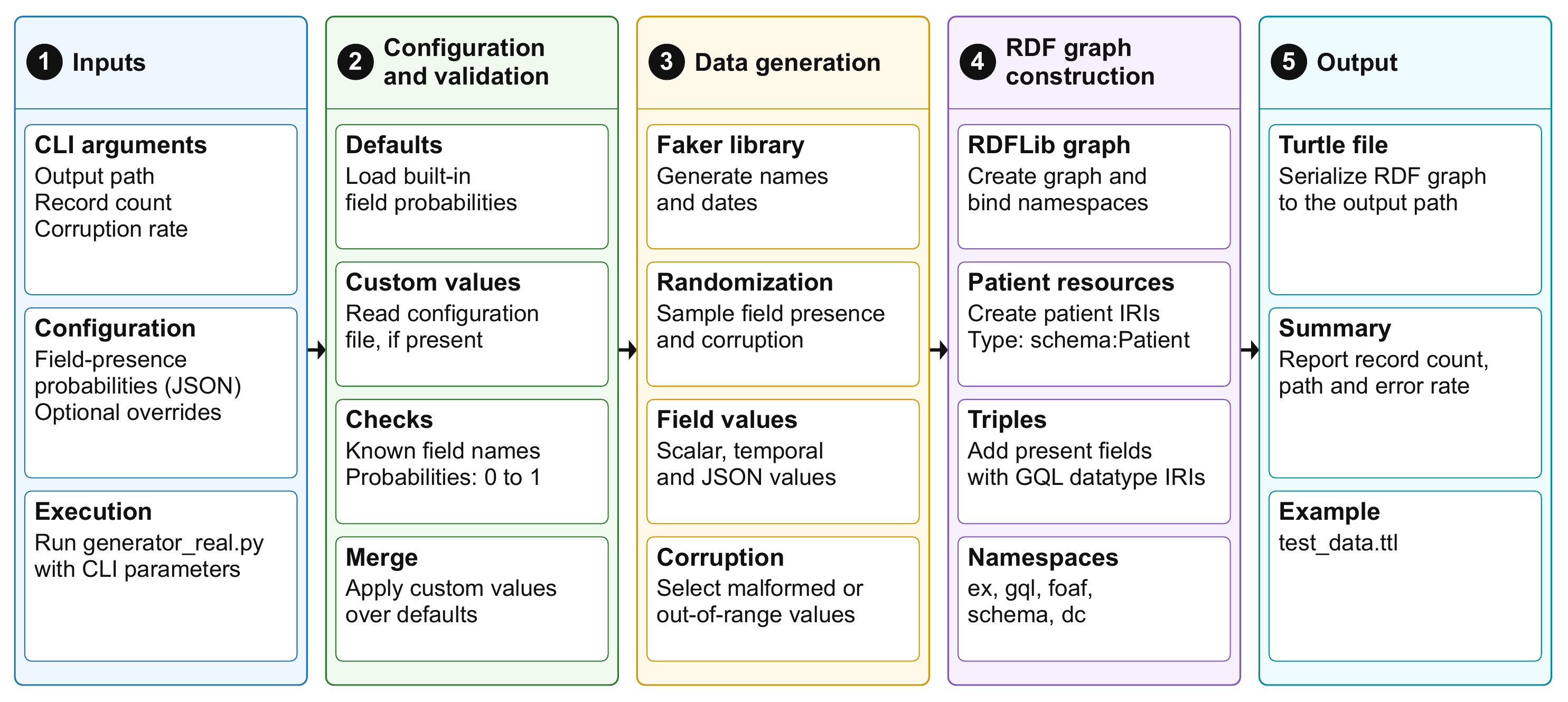}
\caption{Synthetic RDF data generation workflow in the reference implementation.}
\label{fig:generation-workflow}
\end{figure}

The generator creates synthetic patient records with \texttt{rdflib} and \texttt{Faker}; its workflow is shown in Figure~\ref{fig:generation-workflow}. It writes Turtle triples for resources of type \texttt{schema:Patient}, assigns GQL datatype IRIs directly to literal values, and supports configurable field-presence probabilities. Faker supplies names and dates, without modeling clinical distributions or correlations. A command-line error-rate parameter selects corruption candidates, including out-of-range integers, malformed dates and booleans, wrong-length binary strings, JSON objects in place of arrays or vectors, and nonempty unit records. Some candidates are accepted by the validator, such as uppercase \texttt{TRUE}.

In Figure~\ref{fig:generation-workflow}, \stepbadge{1} command-line parameters
select the output, record count, error rate, and probability configuration.
The generator then \stepbadge{2} merges field probabilities, \stepbadge{3}
generates field values and corruption candidates, \stepbadge{4} adds them
to the \texttt{rdflib} graph, and \stepbadge{5} serializes the graph as
Turtle.

\begin{figure}[htbp]
\centering
\includegraphics[width=\linewidth]{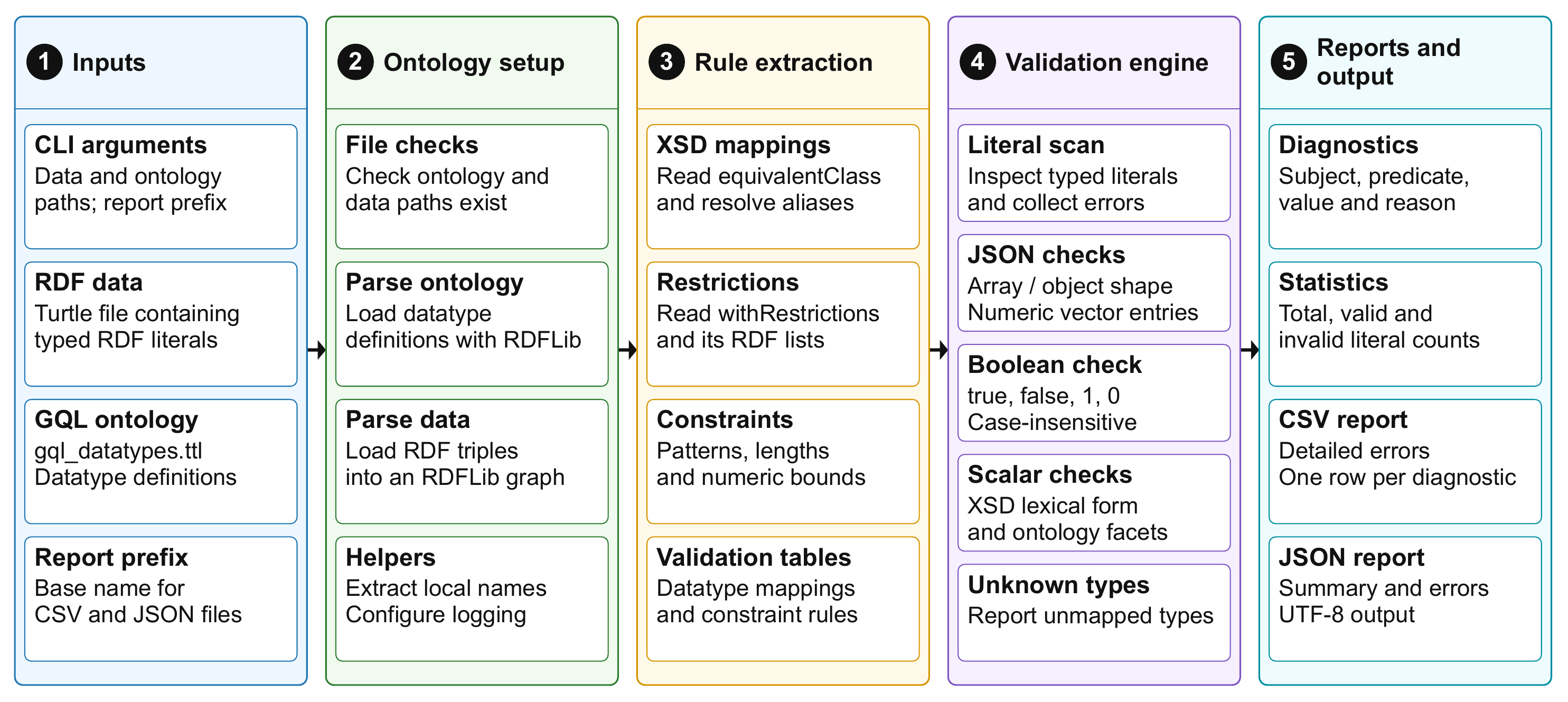}
\caption{Validation workflow for GQL-typed RDF literals in the reference implementation.}
\label{fig:validation-workflow}
\end{figure}

The validator loads the ontology and data graph at runtime, as summarized in Figure~\ref{fig:validation-workflow}. It derives XSD validation bases from \texttt{owl:equivalentClass}, follows aliases whose targets already have an XSD mapping, extracts patterns, bounds, and length constraints from \texttt{owl:withRestrictions}, and validates each typed literal. Validation errors are printed and exported as CSV and JSON reports containing the subject, predicate, value, original GQL datatype, base datatype, and error reason. Thus, the implementation should be read as an executable validation profile for the vocabulary, while the vocabulary itself remains the main contribution of the paper.

In Figure~\ref{fig:validation-workflow}, \stepbadge{1} the validator receives
the data, ontology, and report paths; \stepbadge{2} parses the ontology;
\stepbadge{3} extracts XSD mappings, aliases, and restrictions; and
\stepbadge{4} checks each typed literal. Finally, \stepbadge{5} it prints
diagnostics and writes CSV and JSON reports.
Both tools are standalone Python command-line programs using RDFLib for graph parsing and serialization. Validation is application code, not a registered RDFLib datatype or SHACL extension. Neither workflow executes GQL queries.

\subsection{Evaluation}

We measured execution time and run-to-run variation using datasets of 10,000--50,000
patient records in steps of 5,000. Each size was tested at 10\%, 45\%, and
90\% error-injection rates with ten repetitions, producing 270 runs and
8.1 million records. Field-presence probabilities determine the datatype mix:
age and health-plan flag 0.90; birth date and gender 0.85; weight and identifier
0.80; height 0.75; admission datetime 0.70; prescription 0.65; arrival time
0.60; duration 0.50; salary 0.40; vector and unit record 0.30; unknown type
0.10. Names are mandatory. The test
system had an Intel Core i7-1360P, 62~GiB of memory, Ubuntu 22.04, Python
3.10.12, RDFLib 7.6.0, and Faker 37.0.2. We measured generation and validation
time, throughput, and valid and invalid typed literals~\cite{Szalaj2026}.

Across all runs, the validator evaluated 84,234,831 typed literals, or
approximately 10.40 literals per generated record. Table~\ref{tab:evaluation-errors}
summarizes the validation outcomes. The configured error rate is an
independent corruption probability applied to eligible generated fields; it
is not a target percentage of all RDF literals. The denominator also includes
mandatory and non-corrupted literals, while the optional unknown-datatype
field has its own presence probability. Consequently, the observed invalid
literal rates are lower than the configured injection rates, but increase
monotonically from 8.94\% to 72.79\%.

\begin{table}[htbp]
\caption{Validation outcomes aggregated by configured error-injection rate.}
\label{tab:evaluation-errors}
\centering
\small
\begin{tabular}{rrrrr}
\toprule
Configured & Evaluated & Valid & Invalid & Invalid \\
rate & literals & literals & literals & rate \\
\midrule
10\% & 28,081,282 & 25,570,969 & 2,510,313 & 8.94\% \\
45\% & 28,079,569 & 17,724,368 & 10,355,201 & 36.88\% \\
90\% & 28,073,980 & 7,640,268 & 20,433,712 & 72.79\% \\
\bottomrule
\end{tabular}
\end{table}

\begin{figure}[htbp]
\centering
\includegraphics[width=\linewidth]{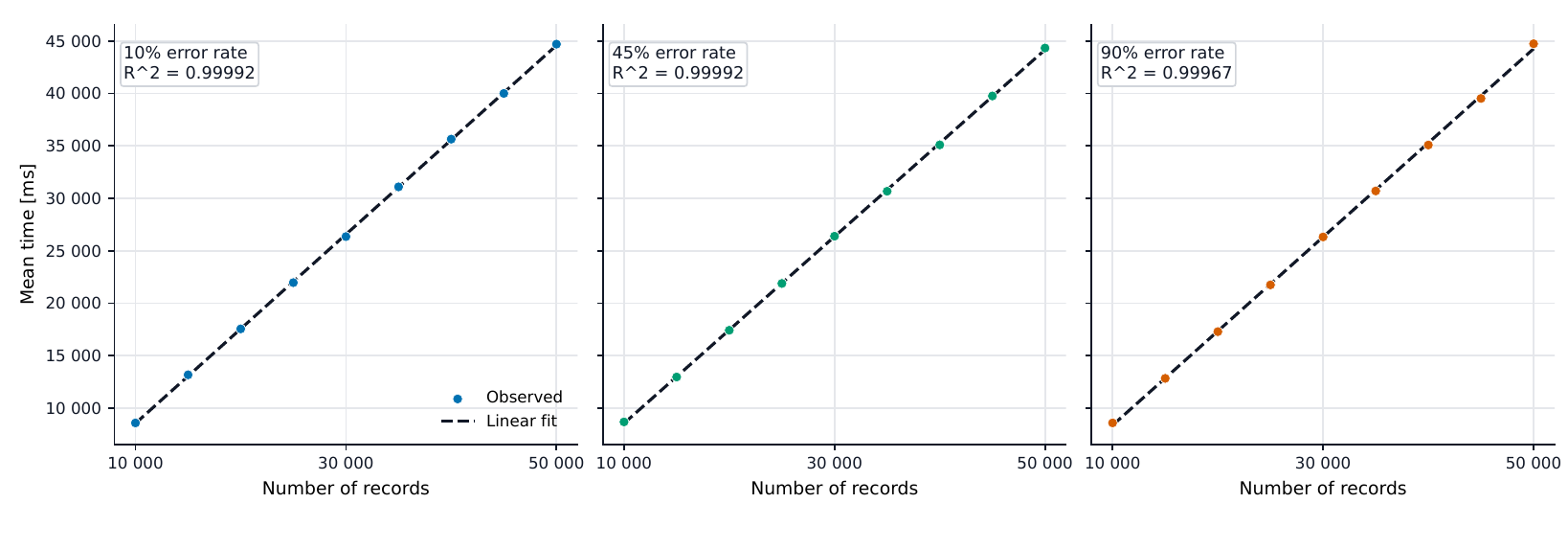}
\caption{Mean generation time and linear fit as a function of the number of
generated records.}
\label{fig:generator-linear-fit}
\end{figure}

\begin{figure}[htbp]
\centering
\includegraphics[width=0.88\linewidth]{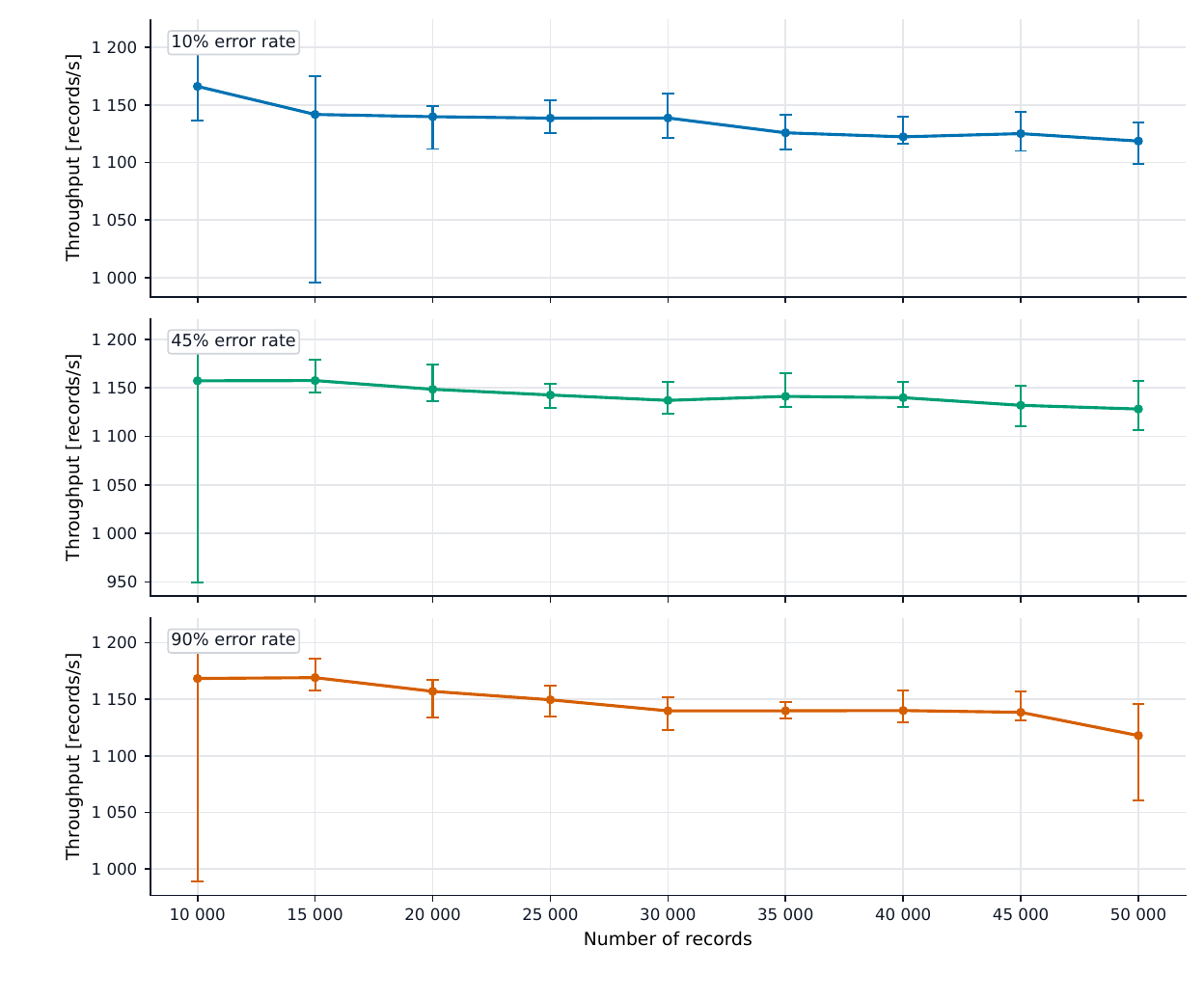}
\caption{Generator throughput by dataset size and configured error rate.
Error bars show the minimum--maximum range over ten runs.}
\label{fig:generator-throughput}
\end{figure}

Figure~\ref{fig:generator-linear-fit} summarizes generation time, while
Figure~\ref{fig:generator-throughput} shows its throughput. Mean
generation time increased from
8.58--8.68~s for 10,000 records to 44.33--44.74~s for 50,000 records.
The differences between the mean and median were small for most
configurations, and the min--max ranges were narrow apart from isolated
runs. The linear models in Figure~\ref{fig:generator-linear-fit} explain more
than 99.96\% of the variation in the mean times
(\(R^2=0.99992\), \(0.99992\), and \(0.99967\) for the three error rates).
Their slopes correspond to approximately 9.01, 8.91, and 8.98~s per
additional 10,000 records. Thus, changing the error-injection rate has little
effect on generation cost.

Generator throughput remains nearly constant across the tested workloads.
The mean values over all dataset sizes were 1,135.1, 1,142.6, and
1,146.8 records/s for the 10\%, 45\%, and 90\% configurations,
respectively. The small differences between these values show that producing
an invalid lexical candidate is not materially more expensive than producing
a valid one. The modest decrease visible for some larger datasets does not
alter the linear scaling trend.

\FloatBarrier

\begin{figure}[htbp]
\centering
\includegraphics[width=0.88\linewidth]{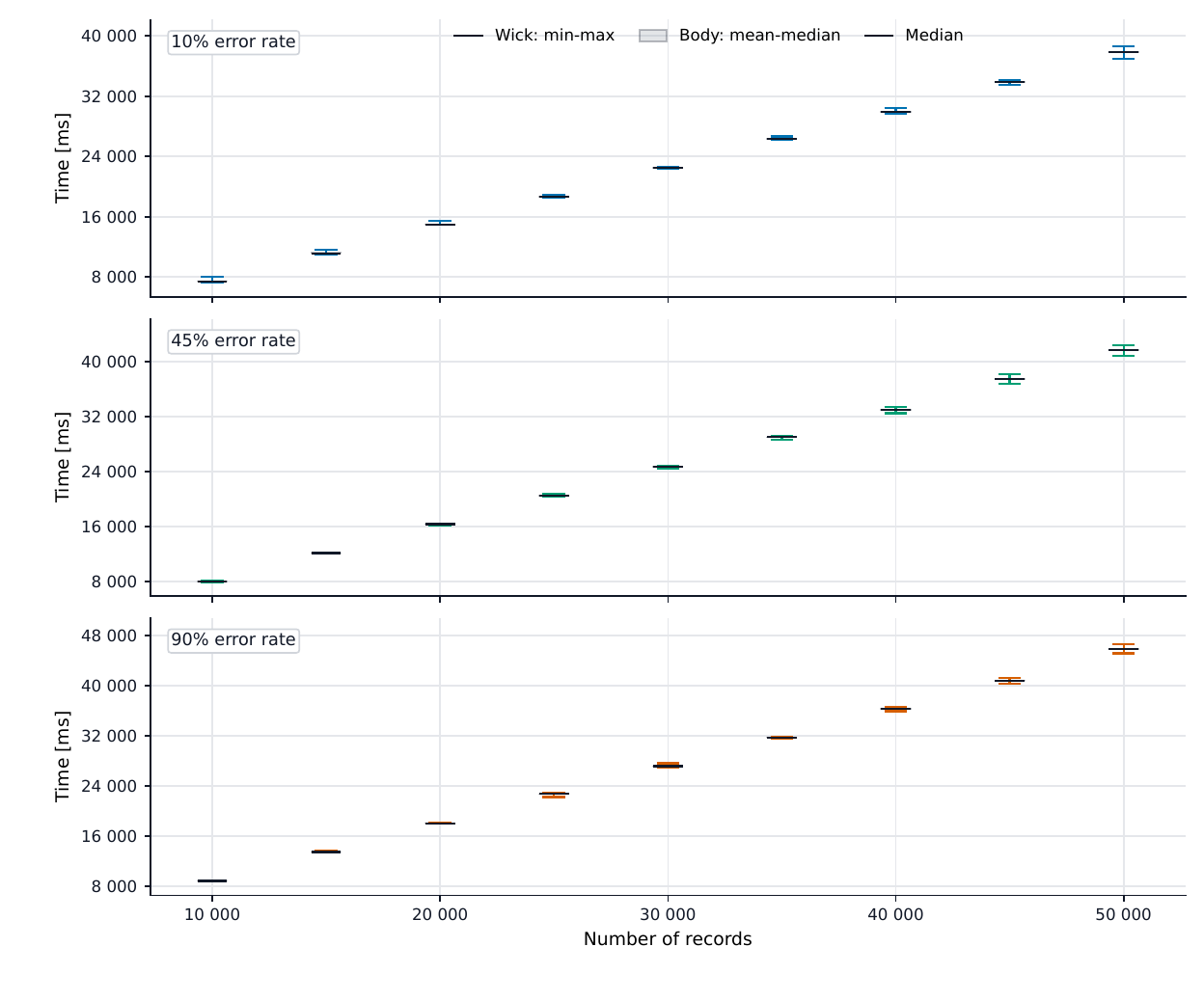}
\caption{Distribution of validation times over ten runs for each dataset
size and configured error rate. Whiskers show the minimum and maximum; the
body spans the mean and median.}
\label{fig:validator-distribution}
\end{figure}

\begin{figure}[htbp]
\centering
\includegraphics[width=\linewidth]{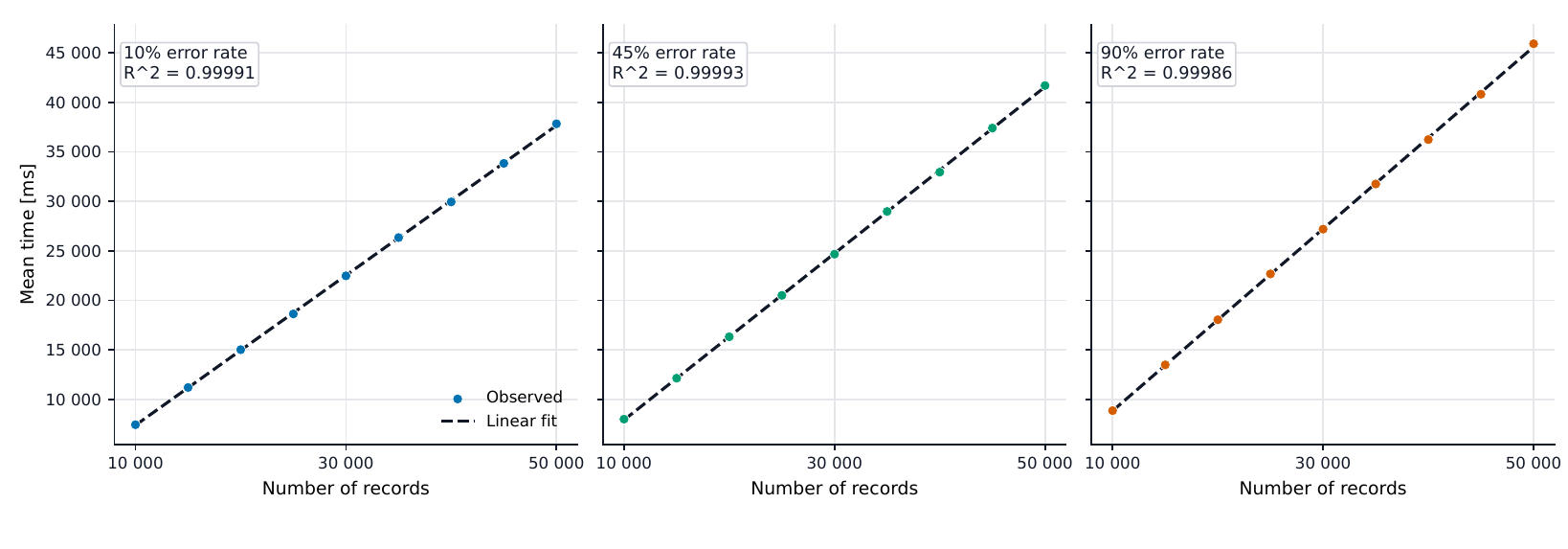}
\caption{Mean validation time and linear fit as a function of the number of
generated records.}
\label{fig:validator-linear-fit}
\end{figure}

\begin{figure}[htbp]
\centering
\includegraphics[width=0.88\linewidth]{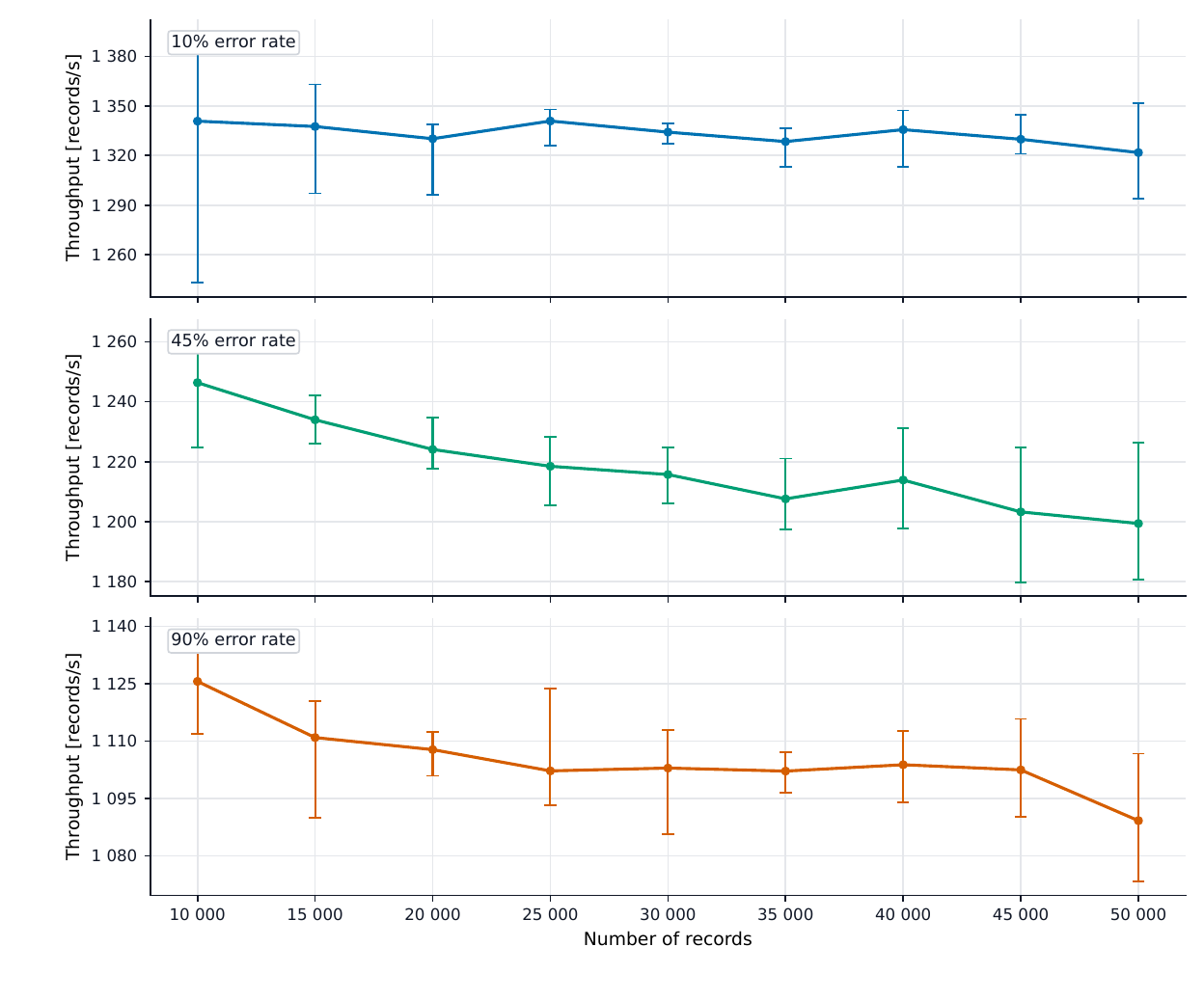}
\caption{Validator throughput by dataset size and configured error rate.
Error bars show the minimum--maximum range over ten runs.}
\label{fig:validator-throughput}
\end{figure}

Validation time is shown in Figures~\ref{fig:validator-distribution} and
\ref{fig:validator-linear-fit}, and its throughput is shown in
Figure~\ref{fig:validator-throughput}.
At 10,000 records, mean validation time was 7.46~s at the 10\% error rate,
8.02~s at 45\%, and 8.88~s at 90\%. At 50,000 records, the corresponding
times were 37.83, 41.69, and 45.91~s. The fitted models remain strongly
linear, with \(R^2=0.99991\), \(0.99993\), and \(0.99986\), while their
slopes increase from 7.56 to 8.40 and 9.18~s per additional 10,000 records.

Unlike generation, validation throughput depends on the number of detected
errors. Its mean value decreased from 1,333.3 records/s at the 10\%
configuration to 1,218.1 records/s at 45\% and 1,105.2 records/s at 90\%.
For the largest dataset, the corresponding throughputs were 1,321.8,
1,199.4, and 1,089.2 records/s. This behavior is consistent with the
additional work required to construct diagnostics and materialize error
reports for invalid literals. Nevertheless, throughput within each error
level remains comparatively stable as the dataset grows.

Overall, the experiment demonstrates approximately linear scaling for both
components over the tested range. The generator is insensitive to the
configured error rate, whereas validator cost increases predictably with the
volume of reported errors. These measurements characterize the reference
implementation on a synthetic workload and a single machine; they should not
be interpreted as a benchmark of RDF stores or production property-graph
systems. Generator measurements characterize workload preparation cost. The experiment does not measure datatype-specific validation cost, conformance accuracy, or end-to-end RDF/property-graph interoperability.

\section{Related Work}
\label{sec:related-work}

RDF collections and containers represent ordered values as graph structures,
with different modeling patterns requiring different SPARQL queries. List.MID
benchmarks six such patterns using MIDI event sequences and corresponding
query templates~\cite{merono2019listmid}. In contrast, this work represents a
GQL list, array, record, or vector as one typed literal, avoiding a list
subgraph but not providing List.MID's element-level graph navigation.

Hartig et al.~\cite{hartig2024datatypes,sparqlcdts} represent lists and maps
as RDF literals and extend SPARQL with constructors, functions, aggregation,
and an unfolding operator. Their approach provides generic composite values
and query operations. This vocabulary instead covers the wider GQL datatype
system, preserves specific GQL datatype IRIs, and provides validation rather
than a SPARQL extension.

LINDT uses application-defined RDF datatypes to encode quantities and units
from the Unified Code for Units of Measure (UCUM) in compact literals~\cite{lindt-custom-datatypes}. It defines lexical and value spaces and provides
datatype-aware comparison, arithmetic, and casting in Apache
Jena~\cite{lindt-custom-datatypes}. Both approaches use RDF datatype IRIs,
but LINDT targets measurement computation, whereas this work
models and validates the GQL datatype taxonomy for graph-data exchange.

RDF 1.2 defines \texttt{rdf:JSON} with JSON/I-JSON lexical forms and a value
space containing strings, numbers, lists, maps, Booleans, and
null~\cite{rdf12-concepts}. It is suitable for generic JSON values. Here, JSON
is only the lexical notation, while datatypes such as \texttt{gql:ARRAY},
\texttt{gql:VECTOR}, and \texttt{gql:UNIT\_RECORD} preserve source-type
information and enable checks not inferable from \texttt{rdf:JSON} alone.

\section{Conclusion}
\label{sec:conclusion}

The GQL Datatypes RDF Vocabulary represents selected GQL value datatypes in RDF. It identifies them by IRIs, records hierarchy and aliasing relations, records selected validator-oriented constraining facets, and supports validation of RDF literals that retain GQL datatype IRIs. By using ontology-derived XSD mappings for scalar datatypes and direct JSON lexical forms for constructed datatypes, the reference implementation preserves GQL datatype information without changing the RDF data model.

The vocabulary supports interoperability between GQL-oriented property graph systems and RDF-based knowledge graph systems. It can be used in data exchange pipelines, RDF exports, schema documentation, and validation scenarios. Since datatype IRIs are RDF resources, they can also be referenced from constraint languages such as SHACL to express GQL-related datatype assumptions. Checking the full GQL-specific constraints, however, requires either generated SHACL shapes, SHACL-SPARQL constraints, entailment-aware validation, or the reference validator described in this paper.

Future work includes three directions. First, the vocabulary should be accompanied by a conformance-oriented test suite that includes positive and negative lexical examples for every datatype. Second, the mapping should be evaluated in end-to-end RDF--property graph exchange scenarios, including round-tripping experiments and datatype preservation metrics. Third, the vocabulary could be integrated more closely with RDF constraint languages and query systems, for example, by generating SHACL shapes that check GQL-compatible lexical forms and by defining reusable datatype profiles for common implementation targets.

\begin{aideclaration}
During the preparation of this work, the authors used OpenAI GPT-5.6 for content drafting, simulated peer review, content refinement, and formatting assistance. All AI-assisted content was carefully reviewed, revised, and validated by the authors, who take full responsibility for the final content of the publication.
\end{aideclaration}

\bibliography{references}

\end{document}